\documentclass[12pt,a4paper]{article}
\usepackage{psfig}

\begin{document}

\title{Breather lattices as pseudospin glasses}

\author{M. Eleftheriou  and G. P. Tsironis \\
Department of Physics, University of Crete\\ and Foundation for 
Research  and Technology-Hellas, P. O. Box 2208, \\71003 Heraklion, Crete, Greece}

\maketitle
\begin{abstract}
We study the thermodynamics of discrete breathers by transforming
a lattice of weakly coupled nonlinear oscillators 
into an effective Ising pseudospin
model. We introduce a replica ensemble and investigate the effective
system susceptibilities through the replica overlap distribution.  We find that
a transition occurs at a given temperature to a new phase characterized by
a slow decay of the relevant correlation functions.  Comparison of
long time pseudospin correlation functions to maximal replica overlap 
demonstrates that the high temperature phase has glassy-like
properties induced by
short range order found in the system.
\end{abstract}

subj-class:Statistical Mechanics

\maketitle

\section{Introduction}
Recent work in the statistical properties of nonlinear lattices has
shown that the presence of discrete breathers (DB's) can induce metastability
and multiple event times that inhibit regular Boltzmann equilibrium at
least for very long observation times \cite{TA}.
It was noted that this behavior is reminiscent that of spin glasses
where an apparently new low temperature phase is induced with seemingly
nonequilibrium properties. Yet, this analogy between nonlinear lattices
and spin glasses has been loose since the latter
are non dynamic spin systems with some form of quenched disorder
while the former are translationally invariant dynamical systems.
It is the aim of the present work to sharpen somehow this connection
by utilizing some of the basic dynamical
properties of breathers and demonstrate that a reduced pseudospin
model can be constructed retaining essential features of the nonlinear
lattice. Furthermore, it will be demonstrated that this reduced pseudospin 
model has a temperature phase with glassy-like properties. As with the case
of spin glasses, glassiness in breather models will be investigated
by analyzing the properties of an appropriate order parameter both in
time and ensemble domain.

Discrete breathers are localized nonlinear modes that exist almost
generically in a vast variety of nonlinear lattice models under some
general conditions \cite{ST,MACAU,AU}. 
In models with nonlinear on site potentials
that is our primary focus here, it is sufficient to choose weak enough coupling
in order to be able to form them. Once formed, breathers are quite stable and
long-lived. The presence of a large number of DB's in the system, for instance
due to coupling with a heat bath or an external field, renders
the translationally invariant lattice into an effectively disordered one with
localized domains of high energy accumulation ('hot spots') while in other
regions there are only linear or quasilinear 
phonon modes\cite{PEY,TA,RAS,RONC,FAR,LIND,Piazza}.  The lattice
is thus naturally split into regions of high local energy accumulation
as well as regions with low energy accumulation. The spatial extent of the
regions can be varied and depends primarily on the specific breather formed.
For hard on site potentials used in the present work, higher energy breathers
are more localized and can occupy essentially only one site with a very
small amount of energy in nearby sites.

The dynamically disordered picture of a nonlinear lattice presented
previously leads naturally in the construction of a reduced pseudospin
Ising-type model for nonlinear lattices exploiting the natural bimodality that
the presence of breathers induces in the system. We can introduce
a projector ${\cal{P}}_i$ that upon acting to each lattice site $i$, gives the
values $+1$ or $-1$ depending whether the local energy at this site is larger
or smaller respectively than a given threshold energy. This threshold
will be determined dynamically in a way to be explained below and is based
on the energy value necessary for specific types of breathers to form
in the lattice. By construction, ``spin up'' corresponds to breather sites
while ``spin down'' corresponds to phonon sites. Once the dynamic
Pseudospin Ising Model (PIM) is formed we may investigate its thermodynamic
properties and in particular its possible glassy behavior, following
standard techniques and ideas taken from the theory of spin glasses\cite{BY}. 
Even
though our PIM does not explicitly involve either quenched disorder or
competing interactions, it nevertheless has these tendencies build in,
albeit in an effective dynamical fashion. Thermal properties of this model
and their connection to the breather system will be analyzed below. In the
following section we present the construction of the model, we discuss
the relevant Edwards-Anderson order parameter in section III, the model entropy
and the dependence on  initial conditions in section IV and conclude in
section V.

\section{Construction of a pseudospin Ising model}
We consider a chain of coupled nonlinear
oscillators with Hamiltonian:

\begin{equation}
H=\sum_{i=1}^{N}(\frac{p_{i}^{2}}{2} +V(x_{i})+\frac{k}{2}(x_{i}-x_{i+1})^2)
\end{equation}

where $x_{i}$ and $p_{i}$ are the displacement and the momentum of the i-th
site respectively.
The lattice is considered large ($N \gg 1$) and periodic $x_{N+1}=x_{1}$.
We choose as on site potential the nonlinear hard $\phi^{4}$ potential,viz.
$V(x_i )=\frac{{x_i}^2}{2}+\frac{{x_i}^4}{4}$,
while the parameter $k$ determines the strength of the nearest-neighbor
interaction.
When we construct breathers using the anticontinuous limit method \cite{AU}
their frequencies for the assumed hard $\phi^4$ potential are higher
than the upper phonon band edge for
each coupling chosen. In this work we use primarily the
value  $k=0.1$ for the coupling that results in
phonon band limits at frequencies $\omega_{ph}=1$ for the lower and
$\omega_{ph}=1.18$ for the upper one.

Let us focus in the regime where breathers are 
primarily localized on a single site or at most they have an extent of
three sites.  This regime can be easily identified from the exact numerical
procedure we follow for the breather construction and the subsequent evaluation
of their energy\cite{AU}.  For instance, a breather of frequency 
$\omega_{b}=1.224$ for $k=0.1$  corresponds to a dimensionless energy
of $E_{b}=0.3004$ while it is localized
mostly on three sites with  $ 60 \% $ of its energy on the central site. 
As a result, if we set an energy threshold equal to 
$ 60 \% $ of $E_{b}$, i.e. $E_{th}=0.18$, then all sites with energy larger or
equal to $E_{th}$ can be considered as "breather sites" while the rest
will be "phonon sites". The specific energy cutoff selects all single-site
high energy breathers and a large majority breathers with
some small extent.  Given the connection between breather amplitude and
its frequency, with the specific selection we
include all breathers that have frequencies higher than ${\omega}_{b}=1.224$.
We note that  this frequency is only $3.7 \%$ higher than the top of the 
linearized phonon band; as a result, the heuristic introduction of the 
specific 
threshold does not lead to serious error, except for the case of the
very extended breathers near the phonon band.  

\begin{figure}[!h]
\centerline{\hbox{
\psfig{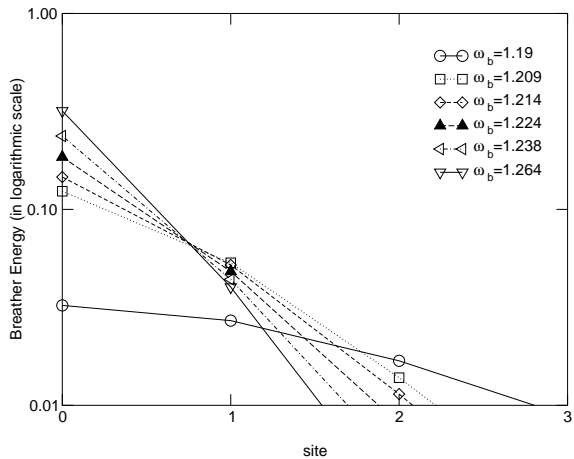}}}
\caption{Local breather energy (in logarithmic scale) as a function of 
site number (central site at 0). We take as a reference the value $0.01$ 
in energy thus neglecting  local energies
below this value. For the three lower frequencies breather is localized
in 5 sites and for the other three frequencies the breather is localized
in 3 sites.
}
\label{fig1}
\end{figure}

Clearly the introduced 
projection underestimates the number of "nonlinear sites" while, additionally,
does not take into account the local lattice coherence in locations where
more than one site breathers exist.  For instance, in the case
of the previous example, while the central site is taken by the projector
to be a "breather site", all its neighbors are considered "phonon sites" by
construction, even though there is a definite coherence between the central
and the nearby sites.  On the other hand, the presence of multibreathers
with energy per site larger than the cutoff is counted correctly.
These issues are depicted in Fig.\ref{fig1} where we present
the energy distribution per site for breathers with different frequencies
for the different sites occupied by the breather. Although
specific results will in general depend on the special choice of a cutoff, 
the general physical behavior should not be sensitive to it.

\vspace{0.8cm}

In order to construct the PIM with local spin $S_{i}$ 
we use the following projector:
\begin{equation}
{\cal{P}}_{i} E_{i} =S_{i}
\end{equation}
where $E_{i}$ is the local energy at site i. The value of the
spin $S_{i}$ is then $S_{i} = +1$ for $E_{i} \geq E_{th}$ and $S_{i} = -1$
for $E_{i} < E_{th}$ . Thus, by construction,  
all positive spins correspond to breather states
while negative spins to linear lattice modes. As the dynamical system evolves 
in time, so does the corresponding Ising model through dynamical spin flipping.
Whenever there is spontaneous energy accumulation, there will be spin up
tendency while when breathers are destroyed the spins will flip down.
The overall spin configuration of PIM will determine the state of the
system and microscopic dynamical events will determine the specific
changes in the spin distribution.  Even though there is no explicit
Ising-like Hamiltonian for PIM, the local dynamics introduces effective
spin-spin interactions that may be quite complex, competing as well as time 
dependent.  As a result, we expect the thermodynamics to be
quite rich with novel aspects.

\section{The order parameter}
The pseudospin Ising model that we established for the hard-$\phi^{4}$
lattice has no direct Hamiltonian representation like the standard Ising
model. Nevertheless spin degrees of freedom change as a result of two factors,
one is the contact to a bath that produces and destroys statistically
breathers while the second is the local breather dynamics itself that can
have similar effects. Although hard to separate the two, the latter tendency
can be seen as a local fluctuating effective spin-spin interaction, similar
in some sense to the competing interaction found in a spin glass (SG).
We note that while in the latter at zero external field there is no 
net magnetization
$m$ in our PIM the value of $m$ is typically nonzero by construction. Although
the transition from negative to positive averaged magnetization marks the
transition from a phonon-dominated to a breather-dominated system, the non-zero
value of $m$ differentiates PIM from a usual SG model.
Nevertheless the presence of breathers at finite temperatures is seen
as a local persistence of the pseudospins in the PIM. This persistence 
introduces short range order in the model and 
affects directly higher correlation spin functions, such as susceptibility.
The relevant quantity that probes these correlations
is the Edwards-Anderson order parameter originally introduced for spin systems
with competition\cite{EA,SK,BY}.  In order to analyze its properties we
will use two approaches, one is a replica representation based on ensembles and
 the second is an analysis in the time domain.
Although in standard statistical mechanical systems ergodicity warranties the
equivalence between phase space averages and time domain averages, this is
not necessary true in more complex systems with some form of competition build
in; the case treated here of the extended nonlinear systems belongs in the
latter category.

\subsection{Replica representation}
In spin glasses the replica approach is used in order to compute the
partition function of the system in the space of variables where the effective
Hamiltonian does not contain any disorder and is translationally invariant
\cite{BY}. Replicas not only help in writing down theoretical relations,
but additionally the replica formalism is very powerful when 
many near equilibrium states of different free energy exist.
Operationally, different replicas can be thought of as clones of the
original system, viz. different statistical realizations
of the same equilibrium system corresponding to different sets of
initial conditions while the overlap between different replicas  
can be thought of as  a measure of the similarity between them \cite{MM}.
The reason for the applicability of replica ideas in the case of breathers
is because the latter, being long-lived metastable states, induce 
state persistence in the nonlinear system and thus non-zero overlap between
different system realizations. As has been already noted in several works, 
the time regime in which standard Gibbsian statistical mechanics is applicable
is not very interesting in most cases.

In order to implement these ideas in the
statistical enumeration of states and study of thermodynamic properties we
introduce system replicas in the following way:   We consider  the
Hamiltonian lattice of $N$ sites (typically $N=200$) and  introduce
initial conditions with random initial velocities following the Gaussian
distribution at temperature $T$, i.e. we start with thermalized velocities but
not thermalized positions.  Clearly the system is out of equilibrium, but not
too far from it.  We then let the system evolve for a time $t_{1}$ that is 
approximately one hundred periods of the cutoff breather period related to the
energy threshold; at the same time we employ numerically the projection 
operation and thus turn the nonlinear system into an effective spin system.  
Longer $t_{1}$-times such as $t_{1}$ approximately $270$ periods does not alter
the results. While in the time window
$t_{1} < t < t_2$ we average the spin values and the outcome defines one
system replica.  This procedure is repeated $n$ times for different initial 
conditions and the resulting ensemble constitutes the replica ensemble at 
temperature $T$.  The typical numbers used are $n=100$ while $t_{1}=400$ and 
$t_{2} =500$ (approximately $78$ and $97$ periods of the cutoff breather period
 respectively);  ensembles up to $n=200$, size of the lattice up to $N=1000$, 
longer times as well as no averaging in the selected time range were
also tested with small changes in the results.   We note that the 
averaging done in the time range $[ t_{1} , t_{2} ]$ simply smooths somehow the
spin distribution of each replica and produces changes only in  cases when 
moving breathers are present. The replica ensemble thus generated can be
thought off as the equivalent equilibrium ensemble for the nonlinear system.
We can now use this ensemble in order to compute the averaged replica 
magnetization defined through

\begin{equation}
\label{magnetization}
m=\frac{1}{\tau ~n ~N} \sum_{t=t_{1}}^{t_{1}+\tau}
\sum_{\alpha=1}^{n} \sum_{i=1}^{N} S_{i}^{\alpha}(t)
=
\frac{1}{nN}\sum_{\alpha=1}^{n} \sum_{i=1}^{N} S_{i}^{\alpha}
\end{equation}

where $\tau$ is the time window $\tau=t_{2}-t_{1}$ that is used for the replica
construction and $S_i^{\alpha}$ is the time independent spin at 
lattice site $i$ belonging to replica $\alpha$.  
We note that even though the time $t_{1}$ is long enough for the
system to reach local thermal equilibrium it is certainly not sufficiently long
for the system to reach true thermodynamic equilibrium.  This is clearly 
observed from the presence of persistent modes in the dynamical lattice or 
the induced short range order in the pseudospin model.  The temperature $T$ 
of the system is defined  through averaged equipartition and the system 
itself is now represented by the collection of the $n$ equivalent time 
independent replicas. The short range order induced by nonlinear localization 
introduces some similarity
between the replicas, even though they were produced in a statistically 
independent fashion. This similarity  between a random pair of
replicas $\alpha$ and $\beta$ can be quantified though the introduction
of the replica overlap:

\begin{equation}
\label{overlap}
q^{\alpha \beta}=\frac{1}{N} \sum_{i=1}^{N} S_{i}^{\alpha} S_{i}^{\beta}
\end{equation}

The quantity $q^{\alpha \beta}$ evaluated for two specific replicas $\alpha$ 
and $\beta$ from the ensemble of $n$ replicas measures the degree of similarity
between the two pseudospin configurations\cite{BY,PA}. Since $n$ is large and
there are $n(n-1)/2$ overlaps one may define some additional quantities that
help quantify the extent of replica similarity; one of those is
the averaged overlap taken over all
replica pairs and being equal to:

\begin{equation}
\label{av-overlap}
q=\frac{2}{n(n-1)} \sum_{\alpha,\beta=1,\alpha \neq \beta}^{n} q^{\alpha \beta}
\end{equation}

Furthermore, the overlap distribution or probability density function 
$P(q^{\alpha \beta})$ that is easily evaluated may provide very detailed
information for the system statistical modes.  In the context of the
Parisi theory\cite{PA}, it can be used in order to evaluate the inverse 
overlap function $x(q)$ that is the cumulant distribution function of $P$, viz.

\begin{equation}
\label{cdf}
x(q)=\int_{-\infty}^{q} dq P(q)
\end{equation}

According to the ideas believed to be true at least in the context of
the mean field theory of spin glasses \cite{SK}, the replica overlap $q(x)$ 
determined as the inverse  function of $x(q)$ defined through Eq. (\ref{cdf}), 
and evaluated at $x=1$, viz. $q(1)$ is identical to the maximum overlap 
$q_{max}$ between all pairs of the replica ensemble.  Furthermore, 
this quantity $q_{max}$ is identified with the Edwards-Anderson order 
parameter for spin glasses that is able to probe into the short range order 
of these systems \cite{PA,BY}. In a regular system, the overlap distribution 
$P(q)$ is a delta function and, as a result, the maximum and averaged replica 
overlaps coincide. In a statistically inhomogeneous system on the other hand, 
the overlap distribution is more complex leading to mean overlap that is 
different from the maximal replica overlap. In these cases, the latter 
determines the tendency of system copies to be similar even though
they were generated through a random procedure. The true thermodynamics of 
these systems is dictated by this property for replica similarity or 
proximity and as a result $q_{max}$ is a pivotal quantity.

\begin{figure}[!h]
\centerline{\hbox{
\psfig{figure=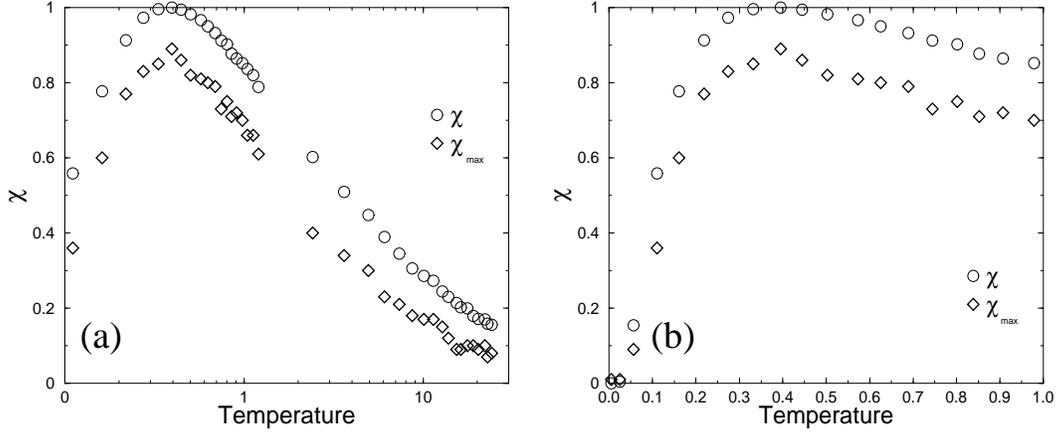,height=6cm}}}        
\caption{Susceptibilities $\chi=1-q$ (circles) and $\chi_{max}=1-q_{max}$ 
(diamonds) as a function of temperature (logarithmic scale in (a)). In (a) 
entire range while in (b) only in the vicinity of the transition point. We 
note the clear difference between $\chi$ and $\chi_{max}$ at high temperatures
where discrete breathers are generated. 
}
\label{fig2}
\end{figure}

We have performed the aforementioned analysis for the $n$ replicas of our 
system and evaluated both the average replica overlap $q$ as well as the 
averaged maximum overlap $q_{max}$ between the replicas. The outcome is shown 
in Fig.\ref{fig2} where $1- q$ and $1- q_{max}$ are plotted as a function of 
the averaged system temperature evaluated through equipartition. The specific 
functional choice for this representation is dictated by the fact that in a 
true magnetical system such as a spin glass, local spin susceptibility 
$\chi$ is related to the overlap $q$ for small fields as  
$k_{B} T \chi =1-q$ in spin glasses \cite{BY}. Thus, $1- q$ and 
$1- q_{max}$ are two measures of the pseudospin susceptibility evaluated
through the averaged and maximal replica overlaps respectively.  The factor
$k_B T$ stems from local application of the fluctuation-dissipation theorem in
standard spin systems; we will not include it in the pseudospin model although
we will still refer to the two quantities derived through the replica overlaps
as "susceptibilities" and use the quantities $\chi$ and $\chi_{max}$ 
respectively to designate them.  After these comments, we can now focus
on the numerical results of Fig.\ref{fig2} and discuss the physics that is 
derived.

The general shape of the PIM susceptibility is similar to that of zero-field
cooled spin glass susceptibility characterized by a reasonably sharp cusp 
at a characteristic temperature $T_g$ \cite{BY}. 
We call hereafter phase I the low temperature regime for 
$T < T_{g}$ while the high temperature regime 
for $T >T_{g}$ we call phase II.
For $ T \ll T_{g}$ both averaged and maximal overlaps give the same 
susceptibility while in the high temperature phase II for 
$T > T_g$ they are clearly different.

While for $T < T_g$ $\chi$ and 
$\chi_{max}$ raise to the maximum in a short temperature range, for 
$T > T_g$ they decay very slowly. At very low temperatures in phase I most 
dynamical states are phonon modes resulting in an ordered state  of down 
spins leading thus to the greatest possible overlap at $T=0$, viz. $q=1$.  As 
the temperature increases more and more breather modes are generated
resulting in spin up states and a subsequent decrease of both the averaged 
and maximal overlaps. The reason for the reduction of the overlap is simple; 
since the breather modes are few and in random lattice locations, mixing 
different random realizations results in smaller overlap.  When the 
temperature reaches the value $T_g \approx 0.38$
the average system magnetization becomes zero and a change in the thermodynamic
behavior occurs.  The onset of persistence due to the abundant breather modes 
results not only in a positive magnetization but also in persistence
in the local pseudomagnetic features.  This is demonstrated by two features of
Phase II in Fig.\ref{fig2}, one being the slow decay as a function of 
temperature of both susceptibilities while the other is the separation of the 
maximal from the averaged susceptibility. 
The slowness of this decay is contrasted to the fast rise on the low 
temperature size of phase I.
 The slow decay of $\chi$ in $T$ 
shows that in phase II the system establishes for each temperature local 
order that is not very sensitive to temperature changes, i.e. the system has 
some "temperature rigidity".  This feature is compatible with the presence of 
a large number of breathers in the system and the fact that random initial 
conditions that do not differ very much result in similar system breather 
content. The second characteristic, viz. that of the 
different decay between $\chi$ and $\chi_{max}$ is a very significant one. 
It demonstrates that the overlap distribution $P(q)$ is not trivial and that
in this regime replicas are now "close" or similar, a feature that is clearly
absent in the phase I.  Physically this proximity of the random replicas in 
phase II is directly attributed to the persistence properties of the breather 
modes.

\begin{figure}[!h]
\centerline{\hbox{
\psfig{figure=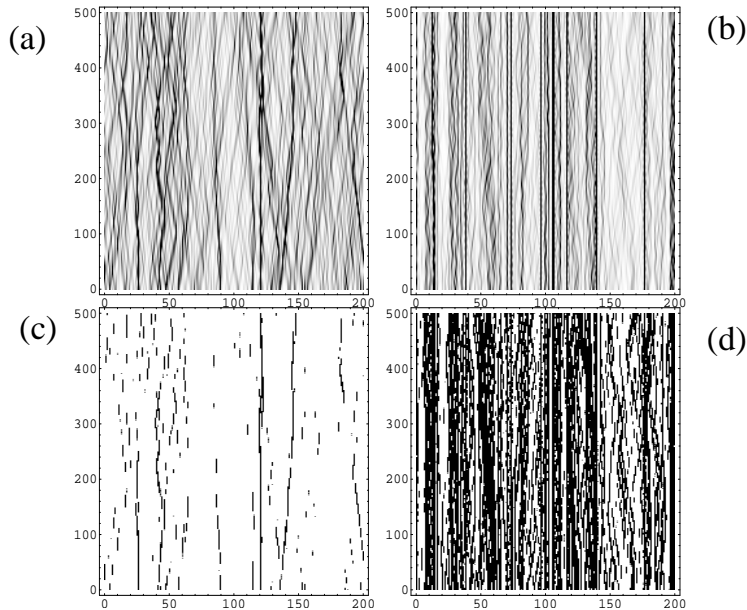,height=8cm}}}        
\caption{Energy density for a system with temperature (a) $T=0.057$ and (b) 
$T=0.533$ respectively; horizontal axis denotes lattice sites while the 
vertical axis denotes time. Black color indicates the more energetic 
regions.  We note that the grey scales in (a) and (b) are normalized as 
intermediate colors between black and white and, as a result, correspond to 
different local energy content in each subfigure.  Graphs 
(c) and (d) are the systems (a) and (b) correspondingly using the pseudospin 
representation. We now note a more precise concentration of high energy 
regions in each case.  Black color designates spin up states.
} 
\label{fig22}
\end{figure}

The contrast between the low and high temperature regime can be seen
easily in an energy density plot and the resulting PIM representation
in each case is shown in Figure (\ref{fig22}).
In Fig. \ref{fig22}(a) the system is at temperature
$T=0.057$ ($k_{B}T<E_{g}$) while in (b) $T=0.533$ ($k_{B}T>E_{g}$). 
The horizontal axis labels the lattice sites, the vertical one is time
while in darker regions the local oscillator energy is higher.
In subfigures \ref{fig22}(a) and \ref{fig22}(b) we show the space-time
evolution of the true local oscillator energy while in (c) and
(d) we show what the system looks like after performing the
pseudospin projection for each corresponding temperature and with
black color labeling the spin up PIM states.
In the first case (Fig. \ref{fig22}(a) and (c)) where breathers cannot occur 
due to low temperature of the system the mean magnetization is negative 
while in the second case (Fig. \ref{fig22}(b) and (d))  breathers have 
formed and the lattice has positive mean magnetization.

Before closing the discussion  on the pseudospin susceptibilities obtained
through the replica method we should make two comments.  Firstly, the 
value of the transition temperature $T_g$ clearly depends on the 
cutoff energy $E_{th}$ of the projector ${\cal{P}}_i$
although not in a trivial way.  While different choices 
of the latter modify $T_g$, the changes  in its precise value are not large. 
Furthermore even though the cutoff $E_{th}$ is determined heuristically, its
physical meaning is very precise and thus model independent. 
We will add more insight into the issue of the physical basis of
the cutoff below.
Secondly, the
function $\chi (T)$ is proportional to $m^2 (T)$ for all temperatures, i.e. the
replica overlap that determines $\chi$ is global and not local.  This feature
which is at variance with usual zero field spin glass behavior stems from the
specific way of construction of the pseudospin model since the latter leads to
non zero averaged magnetization for most temperatures.  Thus, in some sense, 
PIM is related to a SG in the presence of an external field.  We will comment 
further on this point in the conclusions.

\subsection{Time domain statistical mechanics}

In the previous section we investigated thermal PIM properties using a replica
ensemble and found the existence of a high temperature phase with enhanced
replica overlap.  Let us now engage in a complementary, yet more computer 
intensive analysis, performed purely in the time domain and independent of 
the introduction of replicas.  For this purpose we now introduce a time 
averaged local pseudospin correlation function as follows:

\begin{equation}
\label{ssc}
q_{i} (t)=\langle S_{i}(0)S_{i}(t) \rangle=\frac{1}{t_{obs}}\int_{0}^{t_{obs}}S_{i}(t')S_{i}(t'+t) d t'
\end{equation}

where the observation time $t_{obs}$ is taken to be much longer than the
relatively short time scale $t_1$ or $t_2$ and typically equal to
$t_{obs}=10^4$ in time units.  This time-averaged correlation function is
calculated numerically for large lattices of typical size
$N=500$. The true dynamical correlation function is given by:

\begin{equation}
\label{qEA}
{q\prime} =\lim_{t \to \infty}\lim_{N \to \infty}\frac{1}{N}
\sum_{i=1}^{N} \langle S_{i}(0)S_{i}(t) \rangle
= \lim_{t \to \infty}\lim_{N \to \infty} \bar{q}
\end{equation}

where $\bar{q} = 
\sum_{i=1}^{N} \langle q_{i} (t) \rangle / N$
The quantity $q\prime$ measures spin-spin correlations that have not decayed
at a given time $t$ when this time is taken to be infinite, i.e. very long 
compared to all characteristic times of the system.  In the context of the 
Sompolinski theory for SG this quantity is shown to be identical to the 
Edwards-Anderson order parameter, viz. $q\prime = q_{EA}$\cite{SO}.  In 
ordinary mean-field spin glasses at zero external field the non zero value of 
$q_{EA}$ marks the onset of short range order and 
the spin glass phase. One issue of importance is the order of performing the
limits in Eq. (\ref{qEA}). If we take first the time limit, then the system  of
finite size $N$  will have infinite time available to it and thus would
traverse all the parts of phase space. Taking subsequently the thermodynamic
limit would mean that time correlations will be equivalent to a Gibbs average.
If, on the other hand, we take first the limit of $N \to \infty$ then the
system can be trapped in some reduced region of phase space corresponding for
instance in a given distribution of breather states and since its size is
infinite it will not be able to escape from it even in infinite time.
Clearly, in the latter case the correlation function is not a function
of the whole accessible phase space but only of those long lived
parts where the system gets trapped, i.e.  the corresponding correlation 
function is not Gibbssian.

\begin{figure}[!h]
\centerline{\hbox{
\psfig{figure=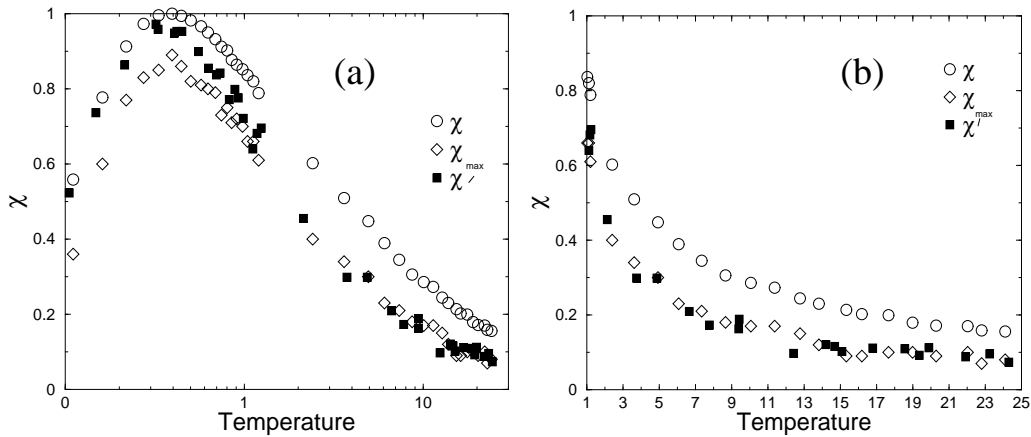,height=6cm}}}      

\caption{Susceptibilities $\chi = 1-q$ (circles), maximal replica 
susceptibility $\chi_{max}=1-q_{max}$ (diamonds) and susceptibility 
$\chi \prime$ obtained from the dynamic overlap $q\prime$ 
($\chi \prime = 1- q\prime$, squares) as a function of temperature 
(in logarithmic scale in (a)). We note in (b) good coincidence between 
$\chi_{max}$ and $\chi \prime$ over the entire high temperature range where
breathers are dominant.
}
\label{fig3}
\end{figure}

The susceptibility $\chi\prime = 1 - q\prime$ of the correlation $q'$ 
evaluated for times of the order $t \approx 10^4$ for system sizes $N =500$ 
is presented in Fig.\ref{fig3} as well as it is compared with the 
susceptibility $\chi_{max}$ determined in the previous section through the 
maximal replica overlap $q_{max}$. 
We observe that there is a good agreement between the two quantities, 
especially in the glassy higher temperature phase II although there are 
clear deviations in the transition region due to increase of 
fluctuations.
This good 
numerical agreement, tested also for other parameter regimes, demonstrates 
that indeed it is the maximal replica overlap that corresponds to the 
long-time system correlation functions. In other words, the thermodynamics 
of phase II is dominated by averages over restricted minima in the system 
free energy. For, if we take into consideration all system copies 
corresponding to different sets of initial conditions  we obtain the average 
magnetization or  equivalently the overlap over all replicas that mixes 
uniformly all the states. The maximum replica overlap on the other hand 
selects only states in close proximity for each temperature and performs the 
averaging only over those states. We observe that the equality 
$q\prime = q_{max} \equiv q_{EA}$ that is expected theoretically in 
mean field spin glass theory holds also in our pseudospin dynamical lattice 
even though the regime identification in the two classes of systems is quite 
distinct. We comment between $\chi \prime$ and $\chi_{max}$ also that the 
deviations observed numerically in the transition region are very likely  
attributed to the numerical time limitations in the evaluation of $q \prime$.

\begin{figure}[!h]
\centerline{\hbox{
\psfig{figure=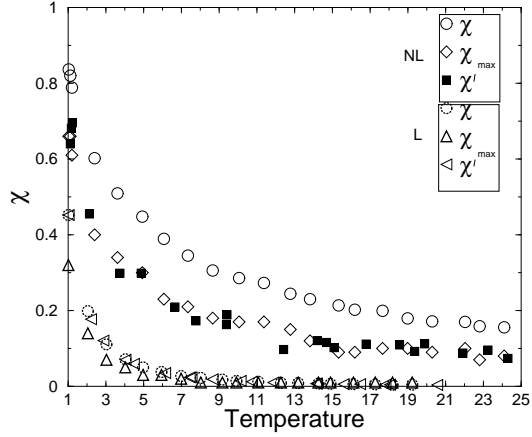,height=6cm}}}      
\caption{Nonlinear case susceptibilities $\chi = 1-q$ (circles), 
maximal replica susceptibility $\chi_{max}=1-q_{max}$ (diamonds) and 
susceptibility $\chi \prime$ obtained from the dynamic overlap $q\prime$ 
($\chi \prime = 1- q\prime$, squares) as in Fig. \ref{fig3}(b). Linear case  
susceptibilities $\chi = 1-q$ (dotted circles),  $\chi_{max}=1-q_{max}$ 
(triangles up), $\chi \prime$ (triangles left). Both sets of susceptibilities 
are plotted as a function of time in the high temperature regime. A sharp 
contrast is seen uppon comparing the three curves of the nonlinear lattice 
with those of the corresponding linear one.
} 
\label{lin}
\end{figure}

We note that the equality between maximal replica overlap and time correlation
function that was found previously does not hold in cases where the dynamical
system is in a regime that does not generate breathers.  We have constructed
a PIM also in the case with a linear on-site potential setting the same energy
threshold that, nevertheless, is highly artificial in this case.  We found 
small differences between the replica averaged,  maximum overlap and spin-spin
correlation function in this case. More specifically, the replica averaged 
and the spin-spin correlation coincide and the maximum overlap has small 
differences from the other two quantities.  In Fig.\ref{lin}  we
show the results for the linear model in the higher temperature phase, as
well as, for comparison, the corresponding values of the nonlinear
model.  We observe a sharp contrast in the scale of the
agreement of the various quantities  in the linear versus the nonlinear
case. We note, however, that in the transition region the
fluctuations are more pronounced, a fact that is expected.

We can make an  analytical estimation of the transition temperature
through the use of the following argument based on energy equipartition
of a single oscillator.  For a single harmonic oscillator with total
energy $E$ and threshold energy $E_{th}$ the transition temperature
can be found
through the median of the Boltzmann distribution $exp ( - \beta E )$ 
determined through the equation
$\int_{E_{th}}^{\infty} e^{-\beta E} dE= \int_{0}^{E_{th}} e^{-\beta E} dE$. 
When this equation holds we have with equal probability a spin up or
a spin down state and as a result the averaged magnetization is zero.
The transition temperature is thus found to be:
\begin{equation}
k_{B} T_c=\frac{E_{th}}{ln2}.
\label{linear_depend}
\end{equation}

The transition temperature for the linear lattice in the low coupling limit  
for energy cutoff $E_{th}=0.18$ is manifested trivially by 
Eq. (\ref{linear_depend}) which gives $T_{c}=0.256$ and agrees quite well 
with the numerical result $T=0.26$. We handle numerically 
several different cutoff values and found that the value of $T_{c}$ is in 
a good agreement with the analytical predictions using 
Eq. (\ref{linear_depend}). The nonlinear case is handled using the same 
argument i.e. the transition temperature is defined through $N_{+}=N_{-}$ 
where $N_{+}=\frac{1}{Z} \int_{E_{th}}^{\infty} e^{-\beta E} 
\frac{d \Gamma}{dE} dE$ and $N_{-}=\frac{1}{Z} \int_{0}^{E_{th}}  e^{-\beta E} 
\frac{d \Gamma}{dE} dE$ and $\frac{d \Gamma}{dE}$ is the density of
states. We found strong deviations between the result resulting
from this one oscillator based calculation 
and the numerical simulations. For the energy 
cutoff $E_{th}=0.18$, which is the one we  mainly use, the transition point 
is through numerical simulation at $T_{g}=0.38$ while the value of 
temperature through the previous argument is at $T=0.288$. This disagreement 
holds for several other different cutoff values as well. We conclude that the 
transition temperature depends on the energy cutoff $E_{th}$ but in the 
nonlinear lattice case it is  clearly dominated by other factors, viz. the 
formation of breathers. Our general conclusion from all the above comments is 
that the equality of the time correlation function $q\prime$ to $q_{max}$ and 
both to the Edwards-Anderson order parameter demonstrates that some form of 
glassiness for the high temperature phase is induced by the nonlinearly 
self-localized breather states.

\section{Entropy of the pseudospin glass state, replica construction and initial conditions}
Using the pseudospin representation for the nonlinear system we can produce
easily an estimate of the part of the system entropy that is related to the 
onset of localized modes. In each temperature the $N$ pseudospins in each 
replica $n$ are partitioned in $N_{+}$ positive and $N_{-} = N - N_{+}$ 
negative ones. The entropy per spin is thus ($k_{B}=1$) :

\begin{equation}
\label{entropy}
S= \ln {N \choose N_{+}} = N\ln N -N_{+}\ln N_{+} -N_{-}\ln N_{-}
\end{equation}

\begin{figure}[!h]
\centerline{\hbox{
\psfig{figure=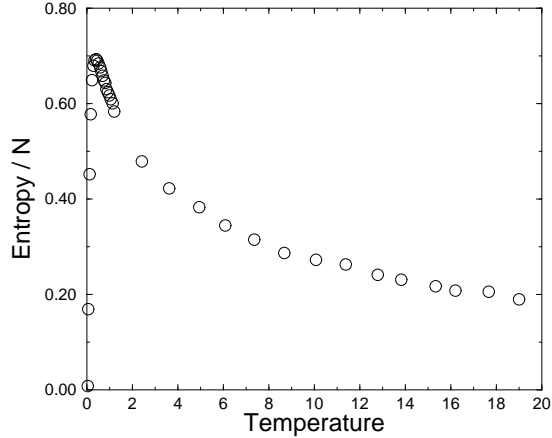,height=6cm}}}
\caption{Numerically evaluated entropy per particle as a function of 
temperature for PIM.}
\label{fig4}
\end{figure}

By construction, at low temperatures, negative spins dominate and the 
entropy is very small; upon increasing temperature, the entropy augments 
and reaches a maximum in the temperature where positive and negative spins 
are equal in number where $S=N \ln 2 \approx  0.698 N$. This transition 
temperature is equal to the one found through the system magnetization and 
susceptibility, viz. $T_g$, since the latter occurs
near the point of zero pseudospin magnetization.  For $T > T_g$ the entropy 
decays slowly as a function of temperature with a rate that is distinctly 
slower than that in phase I; these features are seen in Fig.\ref{fig4} where 
the numerically obtained entropy per spin is plotted as a function of 
temperature.  The entropy reduction at high temperatures is a typical
feature of coupled two level systems in one dimension.  In the 
present model it corresponds to the saturation of the spin up states.  
The form of its occurrence is however different from those in the
standard cases, for, in the present case it does
not happen symmetrically for the low and high temperature phase.
The slow decay in phase II compared to the fast rize in phase I captures
the onset of order in the former case in the form of longer lived
localized states.

We also comment that the calculated entropy  measures the breather entropy 
evaluated over the appropriate free energy minima in  the pseudospin 
representation. All replicas for $T > T_g$ are dominated by nonlinear localized modes and thus the number of up spins does not have a very strong 
temperature dependence. As a result, the calculated entropy is obtained from 
appropriate localized segments of phase space and not the entire phase space.

Let us now focus on the dependence of the results on initial conditions.  The 
role of the latter is critical for the the replica construction
since they enable the system preparation in appropriate phase space states.
Clearly the replica ensemble has to be produced randomly, yet the weight
of nontrivial locally correlated states must be nonzero.  As a result, true
equilibrium initial conditions are not appropriate since they would not 
necessarily produce enough nonlinearly localized states and result 
simply in Gibbs state counting.  This feature of the initial
conditions has been discussed in Ref. (\cite{RAS}).
For our replica preparation we used a Gaussian
distribution over the initial velocities but not the positions of the 
particles. As a result the system initially is away from equilibrium, yet not 
too far from it.  Other initial conditions that are truly nonequilibrium will 
produce a larger weight of the correlated modes in the replica ensemble and as
a result change the specific values obtained for the susceptibilities 
\cite{RAS}.

We have also performed the previous analysis using the Langevin equation 
approach. More specifically, for each temperature $T$ we thermalized the system
using local stochastic noise and dissipation and used this state as the
initial one for the preparation of the replicas as in section III.  We found
the following departures compared to the results presented earlier.  Firstly 
the transition temperature $T_g$ was shifted to a smaller value, i.e. the 
transition to the glass phase occurs earlier. Furthermore, the tail 
distribution for the averaged $\chi$ is not as slow as for the other initial 
conditions while there is now a smaller difference between $q$ and 
$q_{max}$. Also in  the case of Langevin thermalization $q$ and spin-spin 
correlation function coincide together.  These features corroborate
the statements made previously regarding the use of thermalized initial 
conditions. Similar observations have been made in other studies where a 
comparison among different thermalization processes where adopted.

\section{Conclusions}
The analysis of thermal properties of extended nonlinear lattice systems
in their weak coupling limit is a task of paramount importance and difficulty 
due to the formation of nonlinear localized modes.    The latter act as system
impurity modes that nevertheless are self-generated and somehow annealed
in the system. They have considerable resilience to thermal fluctuations, yet
they are produced and destroyed by them.  Their presence modifies the free 
energy landscape of the system and thus the resulting thermodynamics, 
leading to some form of system glassiness.  In order to tackle these issues 
we  first modified our dynamical system and turned it into an effective spin 
system of Ising type where the two spin values where determined from the the
system dynamics through a projection.  Locations with nonlinear energy 
accumulation were mapped to spin up states while the rest to spin down 
states. This reduction of the dynamical system into a spin model enables the 
use of the extended literature of spin glasses where similar issues have been 
addressed.  

In order to handle the presence of long-lived breather modes in the lattice
we used two approaches, one based on system cloning into replicas while the
other in time domain correlation function evaluation.  The replicas where constructed
from a random ensemble of initial velocities and subsequent system evolution
to times that are long for reaching local equilibrium but not long enough for
breather destruction.  The replicas are introduced in order to probe into 
equilibrium thermodynamics while the system is trapped in very long-lived
metastable states.  The basic quantity of interest is the degree of replica
overlap that is directly connected to system susceptibility.  We found that the 
averaged replica overlap leads to a susceptibility that has a relatively sharp
maximum at a given temperature $T_g$ while its values for $T > T_g $ decay slowly
with temperature signifying the presence of a phase with short range order induced
by the breather modes.  The glassy feature of this phase
is demonstrated by the clear difference between the averaged and
the maximal replica overlaps obtained through the calculation of the 
overlap distribution. Use of a complementary approach in the time domain  
shows that the dynamical correlation function evaluated for large systems and 
long times coincides with the maximal replica correlation obtained from the 
replica distribution function. This specific property shows that the high 
temperature phase corresponds to an averaged state over specific sectors of the
phase space related to the presence of nonlinear localized modes and not to the
entire system phase space.

While a reduction of continuous variables into a binary spin variable might
seem at first highly artificial, it is in tune however with the natural 
bimodality that the presence or absence of a breather at each system 
neighborhood introduces. As a result, the pseudospin model we introduced has 
virtues as well as weaknesses.
It is simple enough to be handled with ease as well as to clarify the 
connection of nonlinear lattice physics with spin glass ideas.  While in our 
case of interest there is no quenched disorder, this role is played by the 
spontaneous onset of nonlinear localization, viz. the breathers.  The 
connection of the dynamical system to a spin system enabled us to identify a 
transition to a high temperature phase that has glassy properties.  This 
feature is distinct from ordinary glasses since in the latter it is the lower 
temperature phase that is the glassy one while the phase at higher 
temperature is in liquid form. In our case however, the system at low 
temperatures is linear and thus ordered, leading to a normal state. Glassines 
in the higher temperature phase is obtained due to the spontaneous generation 
of localized modes and the accompanied short range order found in the spin 
system.

One feature of PIM that is not very satisfying is that the formation of the 
specific spin states at each site depends on the local projector with an 
energy threshold. While the concept of the threshold is physically clear, its 
value is determined heuristically
yet it is physically motivated.  Furthermore, its 
specific form produces an approximate spin representation that in the cases 
of single breathers with some extent underestimates the local coherence.  
While the projection enables the construction of a spin model and thus makes 
a contact with the spin glass theory, the PIM is not a standard spin model 
and contains some undesirable features, such as the build in non-zero 
magnetization.  These model deficiencies do not affect however the true value 
of the model to initiate a quantitative study of the glassy breather phase 
with proper concepts and methods.  In a subsequent study we will present the 
physics of this phase as well as the correlation dynamics in the transition 
region independently of the pseudospin model.

We acknowledge partial support from European Union under HPRN-CT-1999-00163
and HPMF-2002-01965 and the Institute of Plasma Physics of the University
of Crete.

\end{document}